\documentclass[useAMS,usenatbib]{mn2e}
\usepackage{graphicx}
\usepackage{epstopdf}
\usepackage{amssymb}
\usepackage{latexsym}
\usepackage {keyval}
\usepackage{amsmath}

\newcommand{\LK} 		{LkH$\alpha$~234}
\newcommand{\sod}		{SO$_2$}

\newcommand{\Tso}		{$^{34}$SO}
\newcommand{\meta}		{CH$_3$OH}
\newcommand{\form}		{H$_2$CO}
\newcommand{\grau}		{$^{\circ}$} 
\newcommand{\SII}		{[S\,{\sevensize II}] }
\newcommand{\kms}		{~km~s$^{-1}$}
\newcommand{\mJy}		{~mJy~beam$^{-1}$}
\newcommand{\msun}	{~M$_{\sun}$}

\newcommand{\cmd}		{~cm$^{-2}$}
\newcommand{\cmt}		{~cm$^{-3}$}
\newcommand{\vlsr}		{$V_{\rm LSR}$}

\newcommand{\eg}		{e.\,g.,}

\newcommand{\DG}		{$\degr$}
\newcommand{\Mm}		{$\pm$}

\title[SMA observations towards the LkH$\alpha$ 234 region]{SMA observations towards the compact, short-lived bipolar water maser outflow in the LkH$\alpha$~234 region}

\author[Girart et al.]{J. M. Girart,$^{1,2}$\thanks{E-mail: girart@ice.cat} J. M. Torrelles,$^3$\thanks{The ICC (UB) is a CSIC-Associated Unit through the ICE} R. Estalella,$^{4}$ S. Curiel,$^{5}$ G. Anglada,$^6$
\newauthor J. F. G\'omez,$^6$ C. Carrasco-Gonz\'alez,$^{7}$ J. Cant\'o,$^5$ L. F. Rodr\'{\i}guez,$^7$
\newauthor N. A. Patel,$^2$ M. A. Trinidad$^8$
\\
$^{1}$Institut de Ci\`encies de l'Espai (CSIC-IEEC), Can Magrans, S/N, 08193 Cerdanyola del Vall\`es, 
Catalonia, Spain\\
$^{2}$Harvard-Smithsonian Center for Astrophysics, 60 Garden Street, Cambridge, MA 02138, USA\\
$^{3}$Institut de Ci\`encies de l'Espai (CSIC-IEEC) and Institut de Ci\`encies del Cosmos (UB-IEEC),\\ ~Mart\'{\i} i Franqu\`{e}s 1, 08028 Barcelona, Catalonia, Spain\\
$^{4}$Departament de F\'{\i}sica Qu\`antica i Astrof\'{\i}sica (formerly Astronomia
i Meteorologia) and Institut de Ci\`{e}ncies del \\
Cosmos (IEEC-UB), Universitat de Barcelona,\\~~Mart\'{\i} i Franqu\`{e}s 1, 08028 Barcelona, Catalonia, Spain\\
$^{5}$Instituto de Astronom\'{\i}a (UNAM), Apartado 70-264, 04510 M\'exico
D. F., M\'exico\\
$^{6}$Instituto de Astrof\'{\i}sica de Andaluc\'{\i}a (CSIC), Apartado 3004, 18080 Granada, Spain\\
$^{7}$Instituto de Radioastronom\'{\i}a y Astrof\'{\i}sica (UNAM), 58089 Morelia, M\'exico\\
$^{8}$Departamento de Astronom\'{\i}a, Universidad de Guanajuato, Apdo. Postal 144, 36000 Guanajuato, M\'exico}

\begin{document}

\date{Accepted XXX. Received 2016 XXXX; in original form 2016 XXX}

\pagerange{\pageref{firstpage}--\pageref{lastpage}} \pubyear{2016}

\maketitle

\label{firstpage}

\begin{abstract}

We present Submillimeter Array (SMA) 1.35~mm subarcsecond angular resolution observations  toward the \LK\ intermediate-mass star-forming region. The dust emission arises from a filamentary structure of $\sim$5~arcsec ($\sim$4500~au) enclosing VLA~1-3 and MM~1, perpendicular to the different outflows detected in the region. 
The most evolved objects are located at the southeastern edge of the dust filamentary structure and the youngest ones at the northeastern edge.
The circumstellar structures around VLA~1, VLA~3, and MM~1  have radii between $\sim$200 and $\sim$375~au and masses in the $\sim$0.08--0.3\msun\ range. The 1.35~mm emission of VLA~2  arises from an unresolved  (r$\la 135$~au) circumstellar disk with a mass of $\sim$0.02\msun. This source is powering a compact ($\sim$4000~au), low radial velocity ($\sim$7\kms) SiO bipolar outflow, close to the plane of the sky. We conclude  that this outflow is the ``large-scale" counterpart of the short-lived, episodic, bipolar outflow observed through H$_2$O masers at much smaller scales ($\sim $180~au), and that has been created by the accumulation of the ejection of several episodic collimated events of material. The circumstellar gas around VLA~2 and VLA~3 is hot ($\sim$130~K) and exhibits velocity gradients  that could trace rotation.   There is a bridge of warm and dense molecular gas connecting VLA~2 and VLA~3.   We discuss the possibility that this bridge could trace a stream of gas between VLA~3 and VLA~2, increasing the accretion rate onto VLA~2 to explain why this source has 
 
an important
outflow activity. 

\end{abstract}

\begin{keywords}
masers -- stars: formation -- ISM: individual objects: LkH$\alpha$~234-VLA~2 -- ISM: jets and outflows -- ISM: molecules
\end{keywords}

\section{Introduction}

Accretion disks and mass-loss processes with the presence of magnetic fields govern the formation of low-mass stars \citep[e.g.][]{mac08, fra14, rao14}.  These mass-loss processes are non-steady, presenting pulsed events that are probably related to episodic increases in the accretion rates 
\citep[e.g.][]{re01, pec10, aud14}. In the case of high-mass protostars, there is increasing evidence that the associated outflows are also non-steady. This is seen, in particular, with the detection through Very Long Baseline Interferometry (VLBI) H$_2$O maser observations of short-lived (tens of years), episodic outflow events, which are interpreted as due to variability in the accretion processes as in the case of low-mass protostars 
\citep[e.g.][]{tor01, san12, tri13, car15}.

\begin{figure*}
\includegraphics[width=17cm]{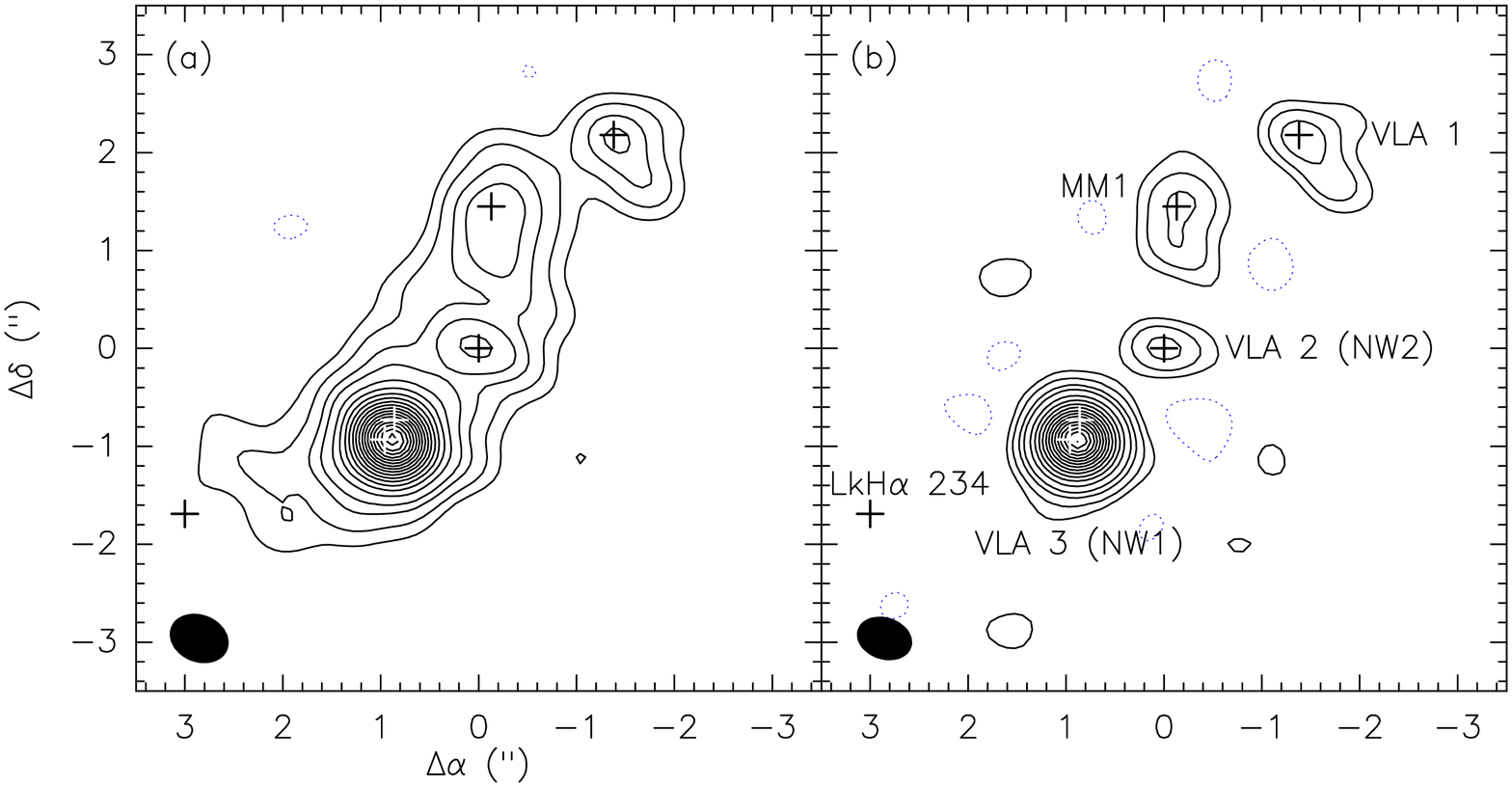}
%\vspace{3.5cm}
\caption{{\it Panel (a):} Contour map of the 1.35~mm continuum emission toward \LK. {\it Panel (b):} same as (a) but using only the visibilities for baselines longer than 100~k$\lambda$.  Contours are $-4$, 4,  8, 14, and from 20 to 170  in steps of 10, times 0.6~\mJy.  The crosses show the position of the YSOs previously reported: \LK, VLA~1, 2, 3A and 3B are taken from \citet[][VLA 3A is the source closer to the dust peak]{tri04} and MM~1 from \citet{fue01}. The black ellipses show the synthesised beams for each map.
}
\label{Fig-cont}
\end{figure*}
\begin{table*}
\caption{1.35 mm continuum sources\label{T1}}
%\begin{tabular}{@{\hspace{0mm}}l@{\hspace{1mm}}r@{\hspace{1mm}}c@{\hspace{1mm}}l}
\begin{tabular}{cccrlrcrr}
\hline
&
\multicolumn{1}{c}{$\alpha$(J2000)} &
\multicolumn{1}{c}{$\delta$(J2000)} &&
\multicolumn{2}{c}{Deconvolved Size$^a$} &
\\
 \cline{2-3}
& \cline{4-5}
& 
\multicolumn{1}{c}{$21^{\rm h} 43^{\rm m}$} &
\multicolumn{1}{c}{$66\degr 6'$} &
\multicolumn{1}{c}{$S_{\nu}$} &
\multicolumn{1}{c}{Major axis, Minor axis} &
\multicolumn{1}{c}{P.A.} &
\multicolumn{1}{c}{$M${\em$^b$}} &
\multicolumn{1}{c}{$n({\rm H_2})$} &
\multicolumn{1}{c}{$N({\rm H_2})$} 
\\
Source & 
\multicolumn{1}{c}{($^{\rm s}$)} &
\multicolumn{1}{c}{($''$)} &
\multicolumn{1}{c}{(mJy)} &
\multicolumn{1}{c}{(arcsec$\times$arcsec)} &
\multicolumn{1}{c}{(\DG)} &
\multicolumn{1}{c}{(M$_{\odot}$)} &
\multicolumn{1}{c}{(\cmt)} &
\multicolumn{1}{c}{(\cmd)} 
\\
\hline
VLA 3	& 6.461 & 55.00 	&139\Mm3	& 0.44\Mm0.02$\times$0.26\Mm0.02	& $-22$\Mm12 
		& 0.25 				& $1.8 \, 10^9$ 		& $5.4\,10^{24}$ \\
VLA 2	& 6.321 & 55.95 	& 12\Mm1	& $\la$0.3$^c$ &
		& 0.023			& $\ga2 \, 10^8$	& $\ga6\,10^{23}$\\
MM 1	& 6.292& 57.26 	& 24\Mm2 	& 0.83\Mm0.15$\times$0.24\Mm0.15	& $-4$\Mm10 
		& 0.09--0.29		& (3--9)$ \, 10^8$		& (1--4)$\,10^{24}$ \\
VLA 1	& 6.084& 58.06 	& 22\Mm2 	& 0.74\Mm0.15$\times$0.23\Mm0.11	&  35\Mm12 
		&  0.08--0.26		& (3--10)$ \, 10^8$		& (1--4)$\,10^{24}$\\
\hline
\end{tabular}
\\
{\em $^a$} Obtained by fitting a Gaussian with the CASA imfit task to each source in the image from Fig.~\ref{Fig-cont}b. The major and minor axis are the full width at half maximum values  of the deconvolved size. \\
{\em $^b$}  Mass estimated from the dust emission assuming: optically thin emission, a gas to dust ratio of 100, a dust opacity per unit of dust mass of 1.1~cm$^2$/g for VLA~2--3 and 0.9~cm$^2/$g for MM~1 and VLA~1, which are the computed values by \citet{Ossenkopf94} at the observed wavelength and for dust particles with thin and thick ices, respectively.   See  Section 3.1 for the value of temperature used for each source. \\
{\em $^c$} The source appears to be unresolved.  The Gaussian fit yields an uncertainty for the major axis of $0.09$~arcsec. We adopt an upper limit which is  three times this uncertainty (i.e., at 3-$\sigma$ level). 
\\
\end{table*}

Very recently, we extended these VLBI water maser studies towards the cluster of intermediate-mass young stellar objects (YSOs) close to the optically visible Herbig Be star LkH$\alpha$234 in the NGC~7129 star-forming region \citep{tor14}. The YSOs in this cluster (including the star LkH$\alpha$234) are grouped within a radius of $\sim$5~arcsec ($\sim$4500~au at a distance of 0.9~kpc; see Kato et al. 2011 and references therein). Five of these YSOs show radio continuum emission: LkH$\alpha$~234, VLA~1, VLA~2, VLA~3A, and VLA~3B \citep{tri04},  all of them, excepting VLA~1, with mid-infrared counterparts \citep{kat11}. From the absence of near-infrared emission in the J, H, and K-bands of VLA~2 and VLA~3A+3B (hereafter VLA~3), but being bright in the $L'$,  $M'$, and mid-infrared bands, \citet{kat11} concluded that these two objects are highly embedded protostars, having also associated H$_2$O maser emission 
\citep[][see also Fig. 1 in Torrelles et al. 2014 showing the spatial distribution of the YSOs in the cluster]{tof95, ume02, tri04, mar05}.
Our multi-epoch VLBI H$_2$O maser observations  \citep{tor14}  revealed a very compact ($\sim$0.2~arcsec, $\sim$180~au), short-lived (kinematic age of $\sim$40~yr),  episodic, and highly collimated water maser bipolar outflow emerging from VLA~2. These results predicted the presence of an accretion disk associated with VLA~2 as well as a relatively compact bipolar molecular outflow when observed through  thermal molecular lines.

In this paper, we present Submillimeter Array (SMA) dust continuum, and thermal molecular line observations at $\sim$1~mm wavelengths carried out towards VLA~2 to detect and characterise the expected disk-YSO-outflow system at scales of a few hundred au (Section~2).  We show the observational results in Section~3,  discussing them in Section~4. The main conclusions of this research are presented in Section~5.

\section{SMA observations}

The SMA observations were carried out on 2014 August 14 and September 1 in the very extended and extended configurations, respectively.  The receiver was tuned to cover the 214.3-218.3 and 226.3-230.3~GHz frequency ranges in the lower (LSB) and upper side band (USB), respectively. The phase center of the telescope was $\alpha$(J2000.0)=$21^{\rm h}43^{\rm m}06\fs50$ and  $\delta$(J2000.0)=$66\degr06\arcmin 55\farcs2$.  We used the flexible SMA  ASIC  correlator, which provides 48 consecutive spectral windows of 104~MHz width.  Most of these spectral windows were set to have a spectral resolution of 1.6~MHz (i.e., 2.13\kms at 230~GHz). However,  some spectral windows, which included several lines (\eg\ \Tso\ 6$_{5}$--5$_{4}$, SO$_2$ 22$_{2,20}$-22$_{1,21}$, CH$_3$OH 5$_{1,4}$--4$_{2 ,2}$, H$_2$S) were set to have a spectral resolution of 0.4~MHz and 0.8~MHz (0.53 and 1.06 \kms, respectively).  Unfortunately these are not the brightest lines, so they are not the best choices to trace the kinematics of the gas.
The gain calibrators were QSOs 1927$+$739 and 2009$+$724. The bandpass calibrator was 3c454.3. The absolute flux scale was determined from observations of MWC349a and Neptune. The flux uncertainty was estimated to be $\sim20$\%.  The data were calibrated using the MIR package\footnote{https://www.cfa.harvard.edu/$\sim$cqi/mircook.html} and imaged using the  MIRIAD software \citep{Sault95}.

Self-calibration (phase only) was performed using the continuum data. The derived gain solutions were applied to the molecular line data.   The continuum maps at 1.35~mm (222.3 GHz) were obtained combining the upper and lower  sidebands, the two array configurations, and using robust weighting of 0.  The resulting synthesised beam was $0\farcs62$$\times$$0\farcs48$  with a position angle P.A.~$\simeq68$\DG.  The rms noise level for the continuum image was $\simeq$0.6~mJy~beam$^{-1}$.  For the molecular line data, the dust continuum emission was subtracted in the visibility space before obtaining the channel maps. The  spectral line maps presented here were obtained by using a robust weighting of 1.0, which yielded a synthesised beam of $0\farcs81\times0\farcs67$ (P.A.~$=73$\DG) for the LSB and of $0\farcs77\times0\farcs63$ (P.A.~$=73$\DG) for the USB.  The figures were created using the GREG package (from the  GILDAS\footnote{GILDAS data reduction package is available at 
http://www.iram.fr/IRAMFR/GILDAS}  software).

\section{Results}

\subsection{Continuum}

Fig.~\ref{Fig-cont}a  shows the 1.35 mm continuum map of the \LK\ region obtained with the SMA.  The emission appears extended and elongated about 5~arcsec (4500~au) in the NW-SE direction, engulfing all known sources but \LK. The overall morphology of the emission agrees with previous, lower angular resolution, 1.35~mm interferometric observations (beam $\sim$1.3~arcsec) by \citet{fue01}.  The observed elongation is also seen, but at much larger scales (up to $\sim$0.3~pc),  by single-dish observation of the 1.35~mm continuum and CO isotopologues \citep{fue01}.

Our  subarcsecond angular resolution 1.35~mm images show four distinct peaks of emission coinciding with sources VLA~1, VLA~2, VLA~3 \citep{tri04}, and MM~1 \citep{fue01}.  MM~1 is the only source not detected at other wavelengths.  For all the mm continuum sources detected with the SMA and reported in this paper there is a systematic position offset of $\sim$0.2~arcsec with respect to the positions measured in the literature at other wavelengths \citep[e.g.,][]{fue01, tri04}.  This value is  comparable to the absolute astrometric uncertainty in a typical SMA observation. In order to correct for the offset, we have used as a reference position the radio continuum position of VLA 2 (the most compact source both at cm and mm wavelengths), 
$\alpha$(J2000) = 21$^{\rm h}$43$^{\rm s}$06.321$^{\rm s}$,  $\delta$(J2000) = 66$^{\circ}$06$'$55.95$''$ given by \citet{tri04}, assuming that the SMA mm continuum peak position of VLA 2 is the same as the VLA 2 position measured at cm wavelengths. This means that we had to apply a small astrometric correction of ($-0\farcs19$, $0\farcs12$) to our SMA data, resulting also in an excellent alignment for all the other sources (see Fig.~\ref{Fig-cont}).  

The total flux detected in the region by the SMA at 1.35~mm is 363\Mm5~mJy.  To compare this value with the one measured by \citet{fue01}  with the IRAM Plateau de Bure interferometer (PdBi) with an angular resolution of $\sim$1.3~arcsec  at 1.22~mm we used the dust continuum spectral index of 2.4 derived by these authors. In this way, we find that the SMA recovers about  90\% of the flux measured by the PdBi.  This is  $\sim$65\% of the total flux measured with the IRAM 30m bolometer \citep{fue01}.

Since there is extended 1.35 mm continuum emission around the four detected sources (VLA 1, VLA 2, VLA 3, and MM1), images have also been made by discarding the visibilities for baselines shorter than 100~k$\lambda$. This allowed us to resolve out the extended filamentary emission and to leave only the emission arising from the four sources. Fig.~\ref{Fig-cont}b shows the resulting map (which yielded a synthesised beam of
$0\farcs58$$\times$$0\farcs43$, P.A.~$\simeq72$\DG), whereas Table~\ref{T1} gives the positions, fluxes, and the deconvolved sizes  by Gaussian fitting of the four sources.  In VLA~3, the brightest source, the emission arises from a region with a radius of $\sim$155~au, suggesting that the dust emission may arise from a circumbinary disk around VLA 3A and VLA 3B. MM~1 and VLA~1 appear to have not only similar fluxes but also similar  sizes. Both sources are elongated, with major and minor semiaxes of $\sim$370 and $\sim$140~au respectively, but with different orientations. VLA~2 appears to be unresolved, with a radius upper limit of $\sim$135~au (Table~\ref{T1}). The total flux of the four compact sources is 197\Mm4~mJy. This leaves a flux that arises from the dust ridge of 166\Mm7~mJy.

The mass, volume and column densities have been computed  from the 1.35 mm dust emission of the four mm continuum sources (Table~\ref{T1}) assuming a constant temperature of 126 and 148~K for VLA~2 and VLA~3, respectively, which is the value derived from fitting the methanol and SO$_2$ lines.  These two lines trace the hot and dense gas at similar scales as the dust (see Tables~\ref{T1}, \ref{T3}, and Section~\ref{mol}). For MM~1 and VLA~1, we assume a temperature range of 30--80~K. The lack of methanol emission in these two sources suggests that it is depleted in the  dust grain mantles \cite[the methanol starts to sublimate at a temperature of $\sim$80~K,][]{Aikawa08}. Adopting a temperature of 22~K, derived from ammonia observations \citep{fue05}, the mass of the dust ridge detected by the SMA is $\sim$7\msun.

\begin{table}
%\begin{minipage}{126mm}
\caption{Molecular lines detected\label{T2}}
\begin{tabular}{l@{\hspace{2.0mm}}c@{\hspace{1.5mm}}crc}
\hline
			& Molecular		& $\nu^a$ & E$_l$ & id. \\
Molecule 	& transition		& (GHz) 	& (K)	& num.$^b$ \\
\hline
SO$_2$
& 16$_{3,13}$--16$_{2,14}$	& 214.68939	& 137.5 & 1 \\
& 17$_{6,12}$--18$_{5,13}$	& 214.72829	& 218.7 & 2 \\
& 22$_{2,20}$-22$_{1,21}$	& 216.64330	& 238.0 & 3 \\
& 14$_{3,11}$--14$_{2,12}$	& 226.30003	& 108.1 & 4 \\
& 11$_{5 7}$--12$_{4, 8}$	& 229.34763	& 111.0 & 5 \\
$^{34}$SO$_2$	
& 4$_{2,2}$--3$_{1,3}$		& 229.85763	& 7.56  & 6 \\
\hline
CH$_3$OH
& 6$_{1,6}$--7$_{2,5}$ v$_t$=1& 215.30220	& 363.5  & 7 \\
& 5$_{1,4}$--4$_{2 ,2}$		& 216.94560	& 45.5  	 & 8 \\
& 6$_{1,5}$--7$_{2,6}$  v$_t$=1& 217.29920	& 363.5 & 9 \\
& 20$_{1,19}$--20$_{0 20}$	& 217.88639	& 497.9 & 10 \\
& 21$_{1, 20}$--21$_{0, 21}$	& 227.09460	& 546.2 & 11 \\
& 16$_{1,16}$--15$_{ 2,13}$	& 227.81465	& 316.3 & 12 \\
& 15$_{4,11}$--16$_{3,13}$	& 229.58907	& 363.4 & 13 \\
&  8$_{-1,8}$--7$_{0,7}$		& 229.75876	& 78.1 	& 14 \\
&  3$_{-2, 2}$--4$_{-1,4}$		& 230.02706	& 28.8 	&  15 \\  
\hline
SO	
&  5$_{5}$--4$_{4}$		& 215.22065	& 33.8 	& 16 \\
$^{34}$SO 
&  6$_{5}$--5$_{4}$		& 215.83992	& 4.0	& 17 \\
H$_2$S	
&  2$_{2,0}$--2$_{1,1}$	& 216.71044	& 73.6  	& 18  \\	
SiO	&  5-4				& 217.10498	& 20.8 	& 19 \\
DCN	& J= 3 - 2		& 217.23853	& 10.4	& 20 \\
H$_2$CO	
&  3$_{0,3}$--2$_{0,2}$	& 218.22219	& 10.5  	& 21 \\
CN	& 2--1 3/2--1/2$^c$	& 226.65956	& 5.4 	&  \\
CN	& 2--1 5/2--3/2$^d$	& 226.87478 	& 5.4 	& \\
HC$_3$N	
& 25--24					& 227.41890	&131.0 	& 22 \\  
C$_4$H    	
& 47/2--45/2$^e$	& 228.53942 	& 316.9	& 23 \\
	\hline
\end{tabular}
\\
%\medskip
\\
{\em $^a$} Frequency from the Cologne Database for Molecular Spectroscopy \citep{Muller01}. \\
{\em $^b$} Identification number shown in Fig~\ref{Fig-spec}. \\
{\em $^c$} The complete transition name is  N=2-1 J=3/2-1/2 F=5/2-3/2  \\
{\em $^d$}  The complete transition name is N=2-1 J=5/2-3/2, F=7/2-5/2.  This line is blended with the J=3/2-1/2 F=5/2-3/2  and  J=5/2-3/2 F=3/2-1/2 lines \\
{\em $^e$}  The complete transition name J=47/2-45/2, $\Omega$=1/2,l=f, v$_7$=1.
\\
\end{table}

\begin{figure}
\includegraphics[width=95mm]{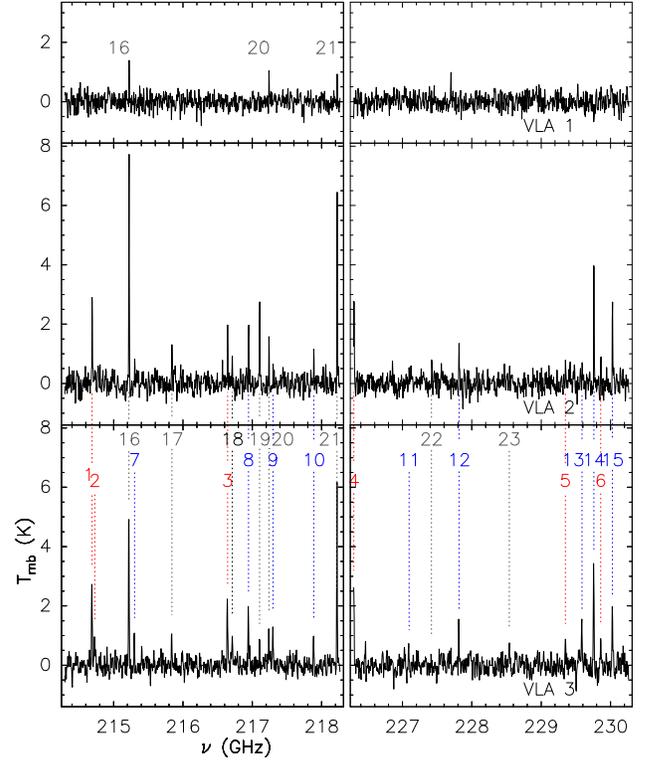}
%\vspace{3.5cm}
\caption{SMA spectra obtained at the position of the dust intensity peak of VLA~1, VLA~2 and VLA~3. Blue and red dotted lines show the \meta\ and \sod\ lines, respectively. Grey dotted lines show other molecular transitions detected. The number is the identification label in Table~\ref{T2}. The conversion from flux density to main beam brightness temperature is $\sim$47~K/Jy at the observed frequencies and angular resolution.
}
\label{Fig-spec}
\end{figure}

\subsection{Molecular lines\label{mol}}

The SMA correlator provides 4~GHz of moderate spectral resolution data (0.5--2.0\kms). Table~\ref{T2} shows the list of detected molecular transitions, including information of  their frequencies and the energy of the lower rotational level.  Regarding their morphology, there are two types of species. On one hand, there are several \sod\ and \meta\ transitions, as well as a C$_4$H vibrational line, that have compact emission and are detected only toward VLA~2 and VLA~3. These transitions appear to have in most cases energy levels above 100~K.  On the other hand, there are several molecular transitions that show extended emission. Fig.~\ref{Fig-spec} shows the rich spectra over the whole observed bandwidth for VLA~2 and VLA~3. For comparison this figure also shows the VLA~1 spectrum.  

 \begin{figure}
\includegraphics[width=70mm]{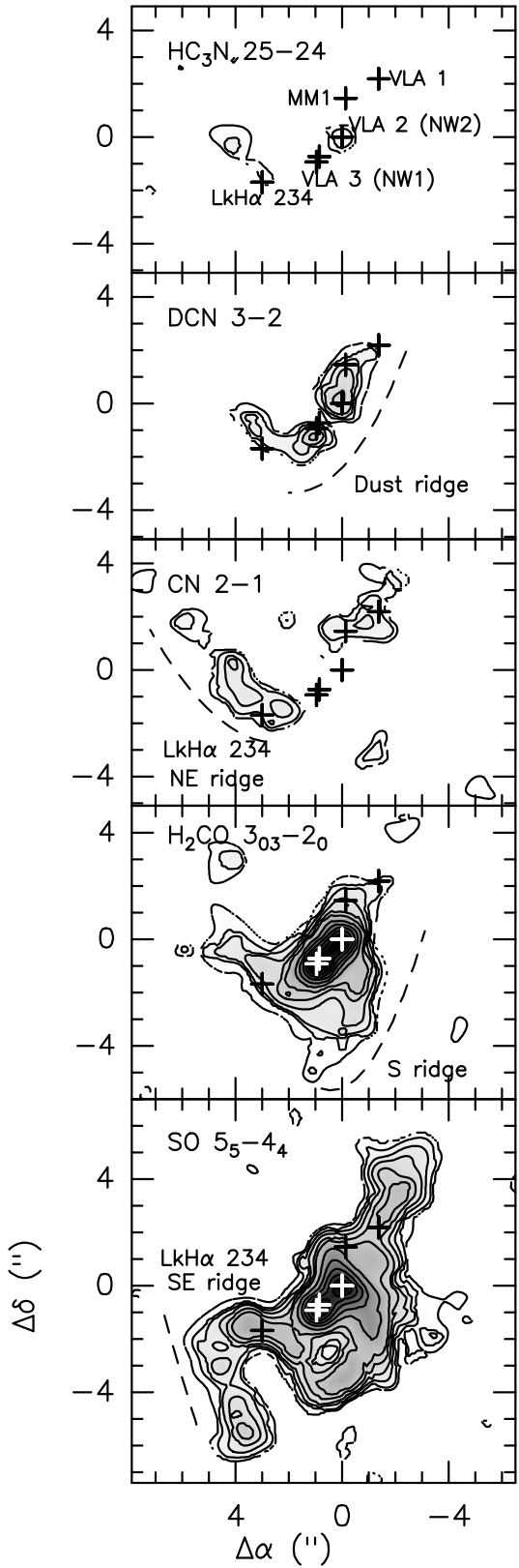}
%\vspace{3.5cm}
\caption{Integrated emission over the \vlsr\ velocity range between $-15.2$ and $-3.8$\kms\ for several molecular transitions that show extended emission. Contours 
are 0.15, 0.3, 0.45, 0.6, 0.8, 1.0, 1.5, 2.0, 2.5, 3.0, 3.5, 4.0~Jy\kms. The crosses mark the positions of the different YSOs in the region.}
\label{Fig-Extnd}
\end{figure}

\subsubsection{Molecular ridges and extended structures around the YSOs}

\begin{figure}
\includegraphics[width=\columnwidth]{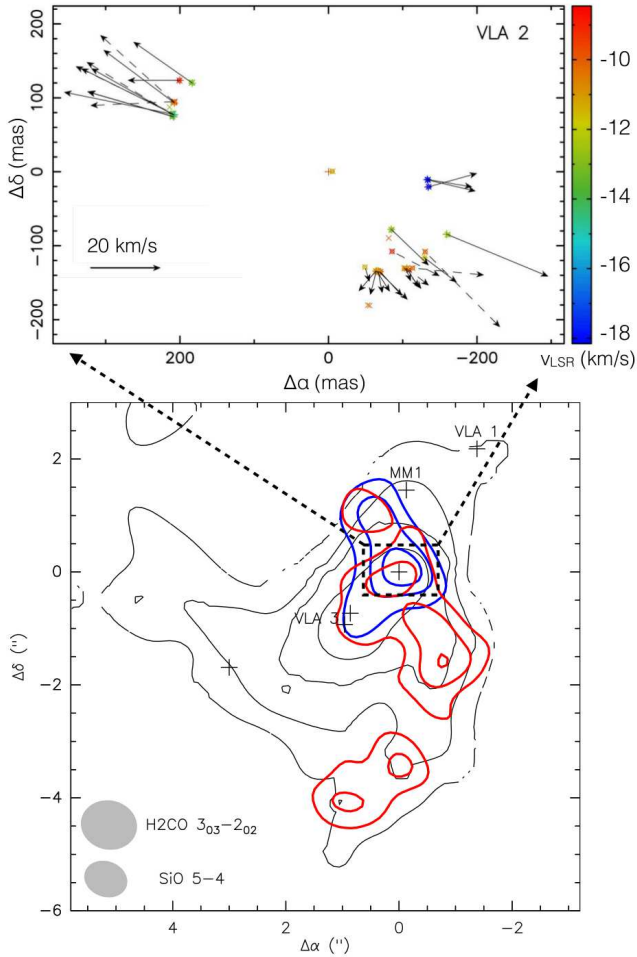}
\caption{
{\it Bottom panel:} Overlap of the \form\ 3$_{0,3}$--2$_{0,2}$ integrated emission (black contours)  with the redshifted (\vlsr\ from $-11.5$ to $-4.5$\kms) and blueshifted (\vlsr\ from $-18.5$ to $-11.5$\kms) SiO 5-4 averaged emission (red and blue contours). The grey ellipses in the bottom part of the panel show the synthesised beam of  these two lines. The crosses mark the different YSOs of the region as in Fig. 1 and 3. \form\ contours are 0.17, 0.56, 1.0 and 2.5~Jy~beam$^{-1}$\kms.
SiO contours are 0.03, 0.06, 0.09 and 0.12~Jy~beam$^{-1}$.
{\it Upper panel}: Close-up of the proper motions of the H$_2$O masers associated with VLA 2 \citep[from][]{tor14}. 
}
\label{Fig-sio}
\end{figure}

Fig.~\ref{Fig-Extnd} shows the integrated emission for the five brightest lines that show extended emission.  These images show a complex chemical behaviour, where each species traces slightly different regions of the \LK\ molecular gas environment. In order to simplify the description, we label in Fig.~\ref{Fig-Extnd} the different molecular structures. Hereafter, we refer to the ``dust ridge'' as the molecular structure that approximately traces the dust emission, as shown in Fig.~\ref{Fig-cont}a. The ``S ridge'' is the curved structure that extends to the south of VLA~2. The ``\LK\ SE ridge'' is the structure seen only in SO to the south-east of \LK.   Finally, the ``\LK\ NE ridge'' is the structure seen to the north-east of \LK.  The length of the different ridges is  $\sim$5~arcsec ($\sim$4500~au), except for the dust ridge that in SO has a length of $\sim$10~arcsec ($\sim$9000~au). The width of the ridges is roughly $\sim$2~arcsec ($\sim$1800 au). Below, we briefly describe the main properties of the spatial distribution of the different molecular transitions:
\begin{itemize}
\item The SO 5$_{5}$--4$_{4}$ line traces all the aforementioned structures 
 except 
the \LK\ NE ridge. There is copious SO emission around all the sources, with the peak emission arising toward VLA~2 and VLA~3.
\item The \form\  line is the second most extended tracer, being detected in all structures 
 except
the \LK\ SE ridge. As in the case of the SO line, the strongest emission arises from VLA~2 and VLA~3.
 \item The CN emission clearly traces the \LK\--NE ridge. There is also fluffy emission around, but not peaking at, VLA~1 and MM~1.
 \item The DCN emission traces mostly the dust ridge although there is a void around VLA~3.  There is also some emission along the  \LK\ SE ridge.
 \item The HC$_3$N 25-24 line is only detected  around VLA~2 and toward the \LK\--NE ridge. 
 \item The SiO 5-4 line emission (Fig.~\ref{Fig-sio}) appears to be detected around VLA~2 and VLA~3 and 
 along the S ridge. However, there are significant differences with respect to the other tracers. 
 There is blue and redshifted emission forming a compact (total length of
$\sim$4~arcsec or $\sim$4000~au), low velocity (outflow velocities of
$\sim$7\kms\ with respect to the VLA~2 systemic velocity,
$\sim$$-11.5$\kms; see next subsection), bipolar outflow in the
northeast-southwest direction (P.A.$\sim20$\DG), that is centred on VLA 2.
The northern blueshifted lobe exhibits some weak redshifted emission. This, and the low radial velocity of the 
 outflow 
suggest that the outflow  axis is close to the plane of the sky (see also Section 4.1.2).   At $\sim$2~arcsec southwest of VLA~2 the redshifted emission bends to the east following the SE ridge. 
\end{itemize}

The radial velocity maps of the molecular gas are shown in the (a) and (b) panels of Fig.~\ref{Fig-vel} for the  \form\ and SO lines, respectively.  There are complex velocity gradients, but the two more interesting ones are: (i) a  mostly smooth gradient along the dust ridge, and (ii) a gradient along the S  ridge (better seen in the \form\ line).   The second order moment map of the SO 5$_{5}$--4$_{4}$ line shows that, except for VLA~2 and VLA~3, the typical velocity dispersion of the molecular gas is $\sim$1.0--1.5\kms (panel c of  Fig.~\ref{Fig-vel}),  which implies a  line width of $\sim$2.4--3.4\kms.  Since the spectral resolution is $\sim$2.3\kms, the line widths are barely resolved.  The  typical velocity dispersion corrected from the channel width is $\la$1.2\kms. The velocity dispersion increases toward VLA~2 and VLA~3 ($\ga 2.0$\kms).

\begin{table}
\caption{ Molecular emission sizes of VLA 2 and VLA 3$^a$\label{T3}}
\begin{tabular}{lcc}
\hline
% &
%\multicolumn{2}{c}{Deconvolved  Size$^a$} \\
Source &
\multicolumn{1}{c}{FWHM} & 
\multicolumn{1}{c}{PA}  \\
\& line &
\multicolumn{1}{c}{(arcsec$\times$arcsec)} & 
\multicolumn{1}{c}{(\DG)} \\
\hline
\multicolumn{3}{c}{\meta\ 8$_{-1,8}$--7$_{0,7}$} \\
\hline
VLA 2(NW2)	& 0.51\Mm0.11$\times$0.37\Mm0.11& $73$\Mm23 \\
VLA 3(NW1)	& 0.66\Mm0.11$\times$0.43\Mm0.13& $-18$\Mm28 \\
\hline
\multicolumn{3}{c}{\sod\ 16$_{3,13}$--16$_{2,14}$} \\
\hline
VLA 2(NW2)	& 0.53\Mm0.22$\times$0.36\Mm0.22& $-52$\Mm70 \\
VLA 3(NW1)	& 0.36\Mm0.12$\times$0.15\Mm0.20& $-47$\Mm44 \\
\hline
\end{tabular}
%\medskip
\\
{\em $^a$}Deconvolved sizes obtained by fitting a Gaussian to each source using the CASA imfit task. We use the integrated emission images from Fig.~\ref{Fig-vel-hc}.\\
\end{table}

\subsubsection{The circumstellar environment around VLA~2 and VLA~3}

The \sod\ and \meta\ lines  (Table~\ref{T2}) are only detected  toward VLA~2 and VLA~3. Most of these lines can only be effectively excited if the gas is very dense and hot (because of their critical densities and energy levels; \eg\ Table~\ref{T2}).

The SO$_2$ and CH$_3$OH lines were used to estimate the temperature of VLA~2 and VLA 3. We performed a rotational temperature diagram analysis\footnote{i.e., we fit ${\rm log} (\int T_{\rm mb} dv /{\rm S}\mu^2 )  \propto E_{\rm u}/T$, where $\int T_{\rm mb} dv$ is the integrated intensity of a line transition, $S$ the so-called line strength of the transition,  $\mu$ the electric dipole moment for the molecule, $E_{\rm u}$  the energy level of the upper transition (in K) and T the gas temperature \citep[e.g., ][]{Girart02}}
using the transitions 1, 2, 3, and 5 of SO$_2$, and the transitions 7 to 15 of CH$_3$OH (Table \ref{T2}). The transition 4 of SO$_2$ was not used because of its location at the edge of the bandpass.  The energies of the transition levels covered a wide range between 30 K and 550~K. The fit of the integrated line intensity of the transitions over $S\mu^2$ versus the energy level  gave us consistent values of the slope for both molecules, for VLA~2 and VLA~3. The value of the slope is inversely proportional to the rotational temperature, assumed to be the same for all the transitions of the same species.  Finally, we obtained rotational temperature values of   $126\pm17$ K and $148\pm21$ K for VLA~2 and VLA~3, respectively (this is the error-weighted average of the SO$_2$ and CH$_3$OH temperatures, see Fig.~\ref{Fig-DiagRot})

 Fig.~\ref{Fig-vel-hc} shows the integrated emission as well as the velocity map of the \sod\ and \meta\ strongest lines. The molecular emission is quite compact, although  around VLA~3 it is slightly elongated along the northwest--southeast direction, with the molecular peaks of both species appearing closer to the continuum centimetre source VLA~3A than to VLA~3B. In order to measure their sizes, we fitted two Gaussians to the integrated emission shown in the aforementioned figure. This fit toward VLA~3 shows that the circumstellar molecular emission has a size and orientation similar to that of the dust emission (Tables~\ref{T1} and~\ref{T3}). The \meta\ and \sod\ emissions trace gas within $\sim$300 and $\sim$160~au (semi-major axes), respectively. Towards VLA 2 the size of the emission is similar for both lines (a semi-major axis of $\sim$230~au). 
 
\begin{figure}
\includegraphics[width=80mm]{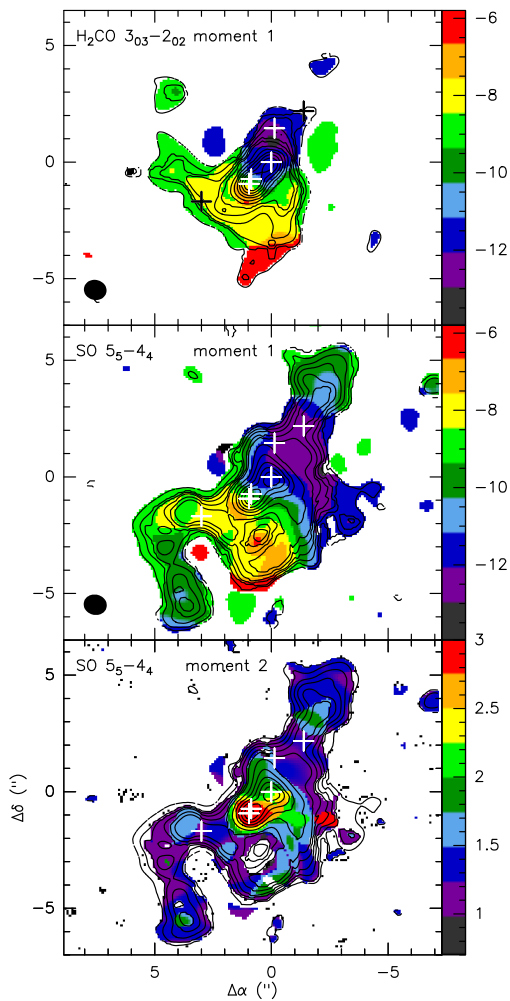}
%\vspace{3.5cm}
\caption{
{\it Panels (a) and (b)}:  Colour image of the first order moment (velocity field) overlapped with the integrated emission of the \form\  3$_{0,3}$--2$_{0,2}$ and SO 5$_{5}$--4$_{4}$ lines, respectively. The colour wedge shows the velocity scale in  \kms. 
{\it Panel (c)}: Second order moment (velocity dispersion) overlapped with the integrated emission of the SO 5$_{5}$--4$_{4}$ line. The wedge indicates the velocity dispersion range in \kms. The crosses indicate the positions of the YSOs as in Figs.~1 and 3.
}
\label{Fig-vel}
\end{figure}

\begin{figure}
\includegraphics[width=\columnwidth]{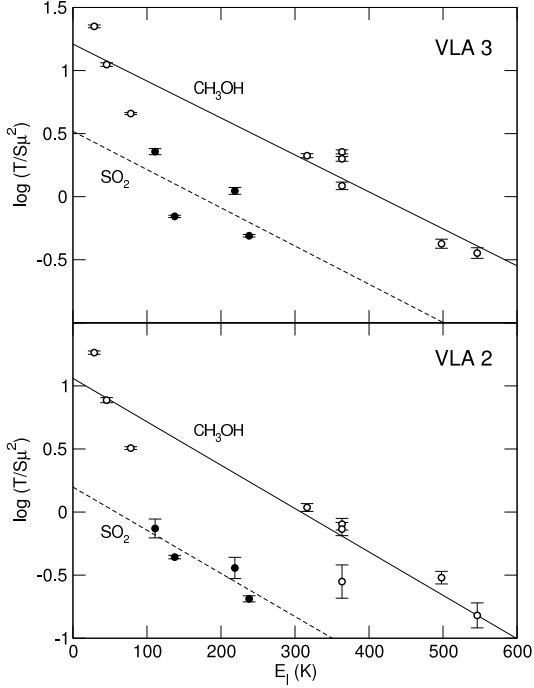}
\caption{
Population diagrams for the SO and the CH$_3$OH. Open and filled circles indicate the observed data for the CH$_3$OH and SO$_2$  transitions, respectively. The error bars include the measurement uncertainty and the calibration uncertainty ($\sim $15\%). The solid (CH$_3$OH) and dashed (SO$_2$) lines show the best fit obtained by assuming optically thin emission (i.e., using the standard population diagram analysis).  For VLA~3 the best fit  is for a temperature of 148\Mm15~K and 144\Mm85~K for the CH$_3$OH and SO$_2$ data, respectively.  For VLA~2 the best fit is 126\Mm13~K (CH$_3$OH) and 127\Mm28~K (SO$_2$).
}
\label{Fig-DiagRot}
\end{figure}

\begin{figure}
\includegraphics[width=80mm]{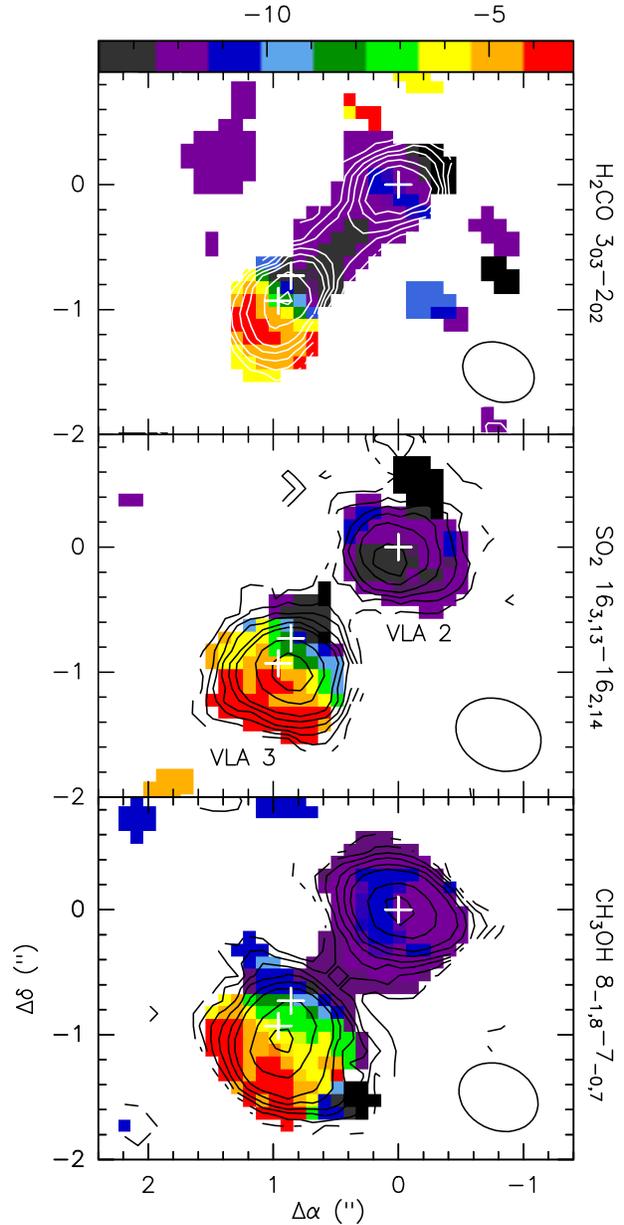}
\caption{
Colour image of the first order moment (velocity field) of the  \form\ 3$_{0,3}$--2$_{0,2}$ (top panel), \sod\ 16$_{3,13}$--16$_{2,14}$ (central panel) and  CH$_3$OH 8$_{-1,8}$--7$_{0,7}$ (bottom panel) overlapped with the contour maps of their integrated emission toward VLA~2 and VLA~3. The wedge on the top indicates the velocity range (in \kms) of the three panels.
}
\label{Fig-vel-hc}
\end{figure}

 The \sod\ and \meta\ lines show a clear velocity gradient, with a total velocity change of $\sim$12\kms, along the major axis of VLA~3 (PA=$-$22\DG; Table~\ref{T1} and Fig.~\ref{Fig-vel-hc}), which is  similar to the direction of the velocity gradient seen at larger scale through other molecular lines (see Fig.~\ref{Fig-vel}).  

Fig.~\ref{Fig-pv} shows  position-velocity plots along the axis that connects VLA 2 and VLA 3 (which is also the dust ridge major axis) for lines typical of warm/hot cores (CH$_3$OH, SO$_2$, H$_2$CO) overlapped with that from the SiO 5-4 line. There are several interesting behaviours in these position--velocity plots: (1) The  center velocity of the molecular emission of the two sources is different, $\sim$$-11.5$\kms\  and $\sim$$-7$\kms\ for VLA~2 and VLA~3, respectively. We note that the $-$11.5\kms\ value for VLA~2 is similar to the mean radial velocity of the H$_2$O masers associated with VLA~2, $-$11.6\kms\ \citep{tor14}.
(2) There is a clear velocity gradient along the major axis of VLA~3: northwest (blueshifted) to southeast (redshifted). In these plots, the emission at the 2-$\sigma$ contour level spans $\sim$20\kms, although the  most blueshifted emission is not detected in the \meta\  line.   Toward VLA~2, the molecular emission at the 2-$\sigma$ contour level spans over a smaller velocity interval, $\sim$12\kms. However, the emission of the  \form\ and \meta\ lines shows only a marginal velocity gradient (in the same direction as in VLA~3).   (3) There is molecular gas between the two sources (as also seen in 
Fig.~\ref{Fig-vel-hc}), with the most blueshifted gas in VLA~3 connected in velocity with a bridge of gas to VLA~2. (4) SiO is detected toward VLA~2 and toward the bridge of gas that connects to VLA~3.

\section{Discussion}

\subsection{The intermediate-mass protostars VLA~2 and VLA~3}

In this section we discuss the properties of these two protostars and the possible interaction between their circumstellar material.

\subsubsection{The masses of VLA~2 and VLA~3}

The dense and warm circumstellar material at scales of 200~au (as traced by the 1.35~mm dust continuum emission) is about ten times more massive in VLA~3 than in VLA~2 ($\sim$0.25 and $\sim$0.023~M$_{\odot}$, respectively; Table~\ref{T1}). VLA~3 has a bolometric luminosity of $\sim$700~L$_{\odot}$ \citep{kat11}. If all the luminosity would come from a single main sequence star, then the stellar mass would be $\sim$6~M$_{\odot}$. However, this is a binary  (VLA~3A+VLA~3B) embedded YSO system, so this mass value is probably an upper limit for each of the two single components (in the case of twin components the mass of each component would be $\sim$5~M$_{\odot}$), even more so considering that part of the luminosity arises from the accretion. The stellar mass for VLA~2 is even more uncertain since the bolometric luminosity is unknown. Yet, its 11.8~$\mu$m flux is  $\sim$2.5 weaker than that of VLA~3  \citep{kat11}, which suggests that VLA~2 is also an intermediate mass star.  In fact, from the observed centimetre continuum emission of VLA 2, \citet{tor14} infer a bolometric luminosity of $\sim$50~L$_{\odot}$ for this object which can be provided by a protostar of a few solar masses. An educated guess of the mass can be also  done by comparing the velocity range of the emission for VLA~2 and VLA~3:  at 2-$\sigma$ level the emission spreads out $\sim$20 and 12~\kms\ for VLA~3 and VLA~2, respectively (see section 3.2.2). If the kinematics is dominated by Keplerian motions 
 (although with the present data we cannot discard the possibility that the motions around VLA~3 are just due to an overlapping of unbound motions of two independent sources, VLA~3A and VLA~3B), 
then the mass ratio between the two sources 
 would be 
$M_{\rm 2}/M_{\rm 3} =  (R_{\rm 3}/R_{\rm 2}) \, (v_{\rm 2}/v_{\rm 3})^2$, where the subindices 2 and 3 refer to VLA 2 and VLA 3. Assuming that the radius of emission, $R$, is approximately the same for the two sources (see section \ref{T3}), then the VLA 2 mass is roughly 40\% smaller than the VLA~3, $\sim$2-4~M$_\odot$.

\subsubsection{The bipolar SiO outflow }

The SiO 5-4 emission appears to originate from two different types of regions. First, there is a low radial velocity ($\sim$7~km~s$^{-1}$) bipolar outflow in the northeast-southwest direction arising from VLA~2.  Fig.~\ref{Fig-sio} shows that the SiO is relatively compact ($\sim$4~arcsec, $\sim$4000~au), specially in the blueshifted  lobe and, more importantly, it is contained within the dense molecular gas structure traced by the \form\ $3_{0,3}$-$2_{0,2}$ line.   As mentioned in Section 3.2.1, the relatively low radial velocities, with the northern blueshifted lobe also having weak redshifted emission, suggest that the outflow is close to the plane of the sky.  There is also redhsifted SiO emission $\sim$4~arcsec south of VLA~3. If this emission is associated with the SiO bipolar outflow, then it would  imply that the redshifted lobe has suffered a deflection of $\sim$50\DG.

We identify the bipolar molecular outflow seen in SiO as the ``extended'' outflow counterpart of the very compact ($\sim$0.2~arcsec, $\sim$180~au), short-lived (kinematic age $\sim$40~yr), episodic,  bipolar outflow observed through H$_2$O maser proper motions also expanding close to the plane of the sky in the northeast-southwest direction with velocities of $\sim$20~km~s$^{-1}$ \citep[see Fig.~\ref{Fig-sio}]{tor14}.  In fact, from the proper motions and radial velocities of the H$_2$O masers reported by Torrelles et al. (2014) we estimate that the outflow axis has an inclination angle of $\sim$15$^{\circ}$ with respect to the plane of the sky.  From the observed spatial extension and radial velocity of the SiO bipolar outflow, we estimate a kinematic age of  $\sim$350~yr for this outflow (after correcting for the inclination angle of $\sim$15$^{\circ}$). We therefore propose that this more extended and  ``older''  outflow seen in SiO has been created by the accumulation of the ejection of several episodic collimated events of material as the one observed through the H$_2$O masers. We note, however, that the position angle of the SiO bipolar outflow is $\sim$206\grau, while the position angle of the very compact H$_2$O maser outflow is $\sim$247\grau\  \citep{tor14}. It is possible that this difference in position angles of $\sim$40\grau\  between the extended SiO outflow and the compact H$_2$O maser outflow is due to several short-lived episodic ejection events, which would have slightly different position angles at their origin.

\subsubsection{A stream of gas  between VLA~3 and VLA~2?}

\begin{figure}
\includegraphics[width=80mm]{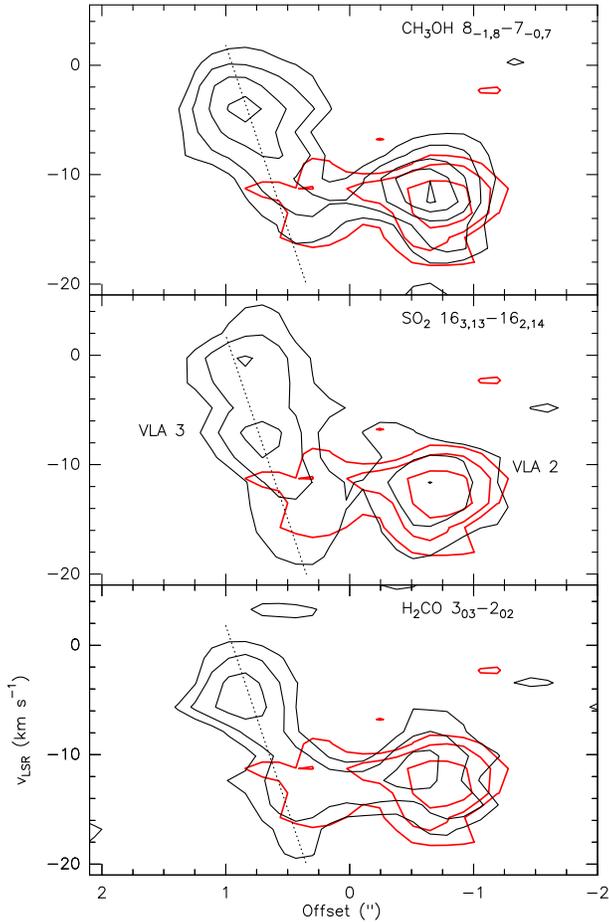}
\caption{
Position-velocity plots along the axis connecting VLA~2 and VLA~3 of the \form\ (bottom panel), \sod\  (middle panel), and \meta\ (top panel) lines (solid contours) overlapped with the SiO 5-4 line (red contours). The  dotted line is shown to approximately mark the velocity gradient in VLA~3. The zero offset position corresponds to the intermediate position between VLA~3 and VLA~2 (see Section 4.1.3). The contour levels are 2, 4, 7, 10 and 13 times the {\em rms} noise of the maps, 23, 20 and 27~\mJy\ for the \meta, \sod\ and \form\ lines, respectively. For the SiO line the contours are 2, 3 and 5 times the {\em rms} noise of the map, 22~\mJy.}
\label{Fig-pv}
\end{figure}

The other type of emission traced by SiO arises at the systemic velocity of the core (see Fig.~\ref{Fig-pv}).  The SiO emission is a signpost of strong shocks, since it can be generated after sputtering in shocks of the Si locked in the dust grains  \citep{Schilke97}. However, in other star-forming regions SiO is detected at velocities close to the systemic velocity of the cloud, with relatively narrow line widths and with no obvious association with the protostellar outflow activity \citep{JS10, NL13}.  It has been proposed that, in  these cases, the SiO can be produced in low velocity shocks ($\la 10$~\kms) if about 10\% of the Si is already in the gas phase or in the icy mantles \citep{NL13}.  Although this can explain the SiO emission at velocities close to the systemic velocity of the core, it is unclear what would be the origin of the initial Si abundance outside of the dust grains.  SiO emission at velocities close to the systemic velocity of the cloud is detected both in VLA 2 and in the bridge of molecular emission connecting VLA~2 and VLA~3 (hereafter ``VLA~3--2 bridge''; see Fig.~\ref{Fig-pv}). Thus, we think that this SiO emission is probably due to the interaction (resulting in low velocity shocks), either between the  VLA 2 and VLA 3 circumstellar gas, or between the VLA 2 circumstellar gas and the dust ridge gas that surrounds the two circumstellar structures. This latter case is less probable, since the VLA~3--2 bridge shows emission from hot core-like molecular lines that are also present in VLA~2 and VLA~3, but not in the dust ridge. Therefore, we favour a  kinematical scenario 
 with a stream of gas between VLA~3 and VLA~2,
producing shocks and releasing SiO. 
 However, with the present data we cannot discern the direction of the stream of gas (either from VLA~3 to VLA~2 or from VLA~2 to VLA~3), particularly considering that VLA~3 is a binary source where we are not able to resolve the kinematics and spatial distribution of the gas contents of each component (3A and 3B). Nevertheless, if the stream of gas is from VLA~3 to VLA~2, this scenario would increase the 
accretion rate onto VLA~2, and could explain why this protostar has a 
 high 
outflow activity.

\subsection{Star formation on the dust ridge}

Here we discuss the apparently organised star formation process taking place in the dust ridge.

\subsubsection{Outflows and the dust ridge orientation\label{S-outflow}}

\begin{figure}
\includegraphics[width=75mm]{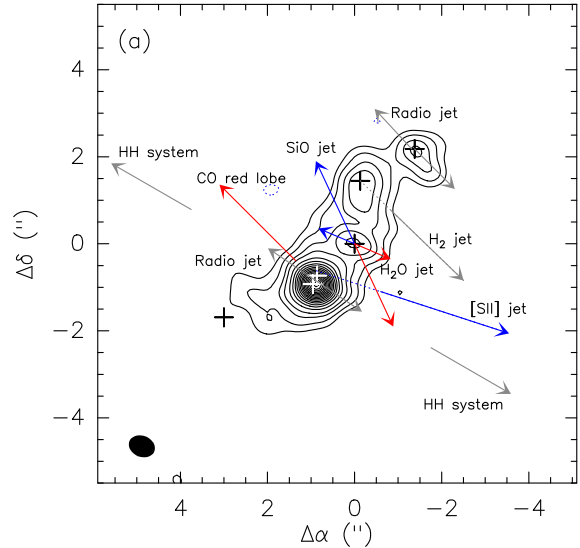}
%\vspace{3.5cm}
\caption{Contour map of the 1.35~mm continuum emission toward \LK\ (same as in Fig.~\ref{Fig-cont}), superposed with the direction of the different outflow signpost detected at different wavelengths (see Sect.~\ref{S-outflow}). Blue and red arrows indicate blue and redshifted gas, whereas grey arrows show outflow direction without velocity information. The crosses mark the different YSOs in the region as in Fig. 1.
}
\label{Fig-cont-out}
\end{figure}

The outflow activity detected in this region is diverse, but all the signposts indicate that there is a preferential direction of the various outflow axes \citep{tri04}.  Besides the aforementioned  H$_2$O maser and SiO outflows associated with VLA~2,  VLA~3A and VLA~3B are radio jets with position angles of  $\sim$55\grau\ and $\sim$57\grau, respectively \citep{tri04}. VLA~1 is also a radio jet with a position angle of  $\sim$45\grau\ \citep{tri04}. There is a large scale ($\sim$1~pc) NE-SW CO bipolar outflow, with the NE red lobe being more extended and brighter than the SW blue lobe \citep{Edwards83, fue01}.  Interestingly, and in spite of having a similar orientation as the SiO bipolar outflow (this paper), the CO outflow has an inverted velocity pattern, suggesting a different powering source. There are two additional jets southwest of the dust ridge. One detected in \SII\, extending $\sim$0.2~pc  with PA$\simeq$252\grau\  \citep{Ray90}.  The other one, in near-IR H$_2$  emission, extending $\sim$0.05~pc with PA$\simeq$226\grau\ \citep{Cabrit97}.    \citet{fue01}  suggest that the \SII\ jet  and the CO NE red lobe are powered by VLA~3, whereas the CO SW blue lobe and the molecular hydrogen jet are powered by MM~1.  Recently, \citet{Oh16} have detected a very high velocity  blueshifted [Fe\,{\sevensize II}] knot, a few arcseconds south of  VLA~3, which appears to be associated with the VLA~3B radio jet.   Finally,  \citet{McGroarty04} detect a chain of HH objects, HH 815-822, centred around the \LK\ region, with an outflow axis at PA$=$60\grau/240\grau\  and extending $\sim$6~pc. However, the driving source of this HH system is unclear. Fig.~\ref{Fig-cont-out} shows the dust emission traced by our SMA observations with the directions of the outflows identified in the region.  It is clear that the overall trend of the activity of the outflows occurs in a direction that is approximately perpendicular to the dust ridge. We think that a magnetic field perpendicular to the dust ridge could produce the general alignment of the different outflows in the region. Although in general there is not a clear correlation between the magnetic field and the outflow direction at core scales, $\sim 10^3$~au \citep{Girart06, Hull13, Zhang14}, there are some cases where they are aligned \citep{Qiu14, Zhang14, Hull14}.
This can be tested with dust polarisation measurements along the dust ridge.

\subsubsection{Organised star formation?}

The most evolved source, \LK, is located at the south-eastern edge of the dust ridge, whereas the two apparently youngest sources, MM~1 and VLA~1, are located at the opposite side of the dust ridge.   Our estimate of these sources being the youngest is based on two main facts. First, they are not detected at mid-IR or shorter wavelengths. Second, the circumstellar gas is as dense and massive as in VLA 2 and VLA 3, but it is significantly colder (there is no indication of hot core tracers as in VLA~2  and VLA~3).  Therefore, these properties suggest an overall evolutionary sequence along the dust ridge, where star formation started in the southeastern  part of the dust ridge and has proceeded with time toward the northwest. We speculate that a compression  wave has continued toward the northwest extreme, once  the star formation started in \LK. This, together with the fact that the outflow activity occurs in a direction perpendicular to the dust ridge suggests that the present star formation activity in the region around  \LK\  is taking place in an organised fashion.

\section{Conclusions}

We report SMA 1.35~mm continuum and spectral line observations  (angular resolution $\sim$0.5-0.8~arcsec) towards the intermediate-mass star-forming region around the optically visible star LkH$\alpha$~234. 

We have detected a dust ridge of $\sim$5~arcsec ($\sim$4500~au) size containing the cluster of YSOs VLA~1, VLA~2, VLA~3A, VLA~3B, and MM1. The mass of the dust ridge is $\sim$7~M$_\odot$ while the circumstellar masses around these YSOs are in the range of $\sim$0.02--0.3~M$_\odot$. We find an evolutionary star formation sequence along the dust ridge, with the most evolved  objects located at the southeastern edge of the ridge and the youngest ones located at the northeastern edge.
The outflow activity of the different YSOs in the region occurs approximately perpendicular to the dust ridge. This overall trend of orientation of the outflows could be explained by the presence of a ``large-scale" magnetic field perpendicular to the dust
ridge.

Our SMA observations reveal towards VLA~2, a compact ($\sim$4~arcsec, $\sim$4000~au), SiO bipolar outflow, moving close to the plane of the sky in the northeast-southwest direction. The kinematic age of the SiO outflow is $\sim$350~yr. We conclude that this outflow is the ``large-scale" counterpart of the much more compact ($\sim$0.2~arcsec, $\sim$180~au), short-lived ($\sim$40~yr), episodic, bipolar H$_2$O maser outflow previously reported with VLBI observations towards VLA~2. We propose that the bipolar outflow seen in SiO has been created by the accumulation of material through repetitive ejections of collimated gas as the one observed through the H$_2$O masers.

There is a bridge of molecular gas connecting VLA~3 and VLA~2 having SiO emission at ambient velocities. We  discuss the possibility that this SiO ambient emission in the VLA~3--2 molecular bridge is due to an interaction of the gas transfer  between VLA~3 and VLA~2.  If the gas transfer occurs from VLA~3 to VLA~2 it could increase the accretion rate onto VLA~2, explaining the observed high outflow activity in this source.

\section*{Acknowledgments}

GA, RE, JMG, JFG, and JMT acknowledge support from MINECO (Spain) AYA2014-57369-C3 grant (co-funded with FEDER funds). JMG also acknowledges support from  the MECD PRX15/00435 grant (Spain) and from the SI CGPS award, ``Magnetic Fields and Massive Star Formation''.  RE also acknowledges MDM-2014-0369 of ICCUB (Unidad de Excelencia `Mar\'{\i}a de Maeztu'). SC acknowledges the support of DGAPA, UNAM, and CONACyT (M\'exico).

\label{lastpage}

\bsp

\end{document}